\pdfoutput=1
\documentclass[conference]{IEEEtran}
\IEEEoverridecommandlockouts

\usepackage{cite}
\usepackage{amsmath,amssymb,amsfonts}
\usepackage{amsthm}
\usepackage{mathtools}
\usepackage{booktabs}
\usepackage{array}
\usepackage{multirow}
\usepackage{graphicx}
\usepackage{xcolor}
\usepackage{url}
\usepackage{tikz}
\usetikzlibrary{positioning,arrows.meta}
\usepackage[hidelinks]{hyperref}

\newtheorem{theorem}{Theorem}
\newtheorem{corollary}[theorem]{Corollary}
\newtheorem{proposition}[theorem]{Proposition}

\newtheorem{remark}{Remark}
\theoremstyle{definition}
\newtheorem{definition}{Definition}

\DeclareMathOperator*{\argmin}{arg\,min}

\newcommand{\inner}[2]{\langle #1, #2 \rangle}
\newcommand{\norm}[1]{\| #1 \|}
\newcommand{\R}{\mathbb{R}}

\newcommand{\Prob}{\mathbb{P}}
\newcommand{\udir}{\hat{u}_{\mathrm{dir}}}
\newcommand{\sg}{\mathrm{sg}}
\newcommand{\mns}[2]{%
  \begin{tabular}[t]{@{}c@{}}#1\\[-3pt]{\tiny $\pm$#2}\end{tabular}%
}

\title{LQ-Rec: Late Quantization for Sequential Recommendation}

\author{%
  \IEEEauthorblockN{Satoshi Koide}
  \IEEEauthorblockA{Toyota Motor North America\\
    \texttt{satoshi.koide@toyota.com}}
  \and
  \IEEEauthorblockN{Yueqi Wang}
  \IEEEauthorblockA{UC San Diego\\
    \texttt{wyq@ucsd.edu}}
  \and
  \IEEEauthorblockN{Julian McAuley}
  \IEEEauthorblockA{UC San Diego\\
    \texttt{jmcauley@ucsd.edu}}
\thanks{\copyright~2026 IEEE. Personal use of this material is permitted. Permission from IEEE must be obtained for all other uses, in any current or future media, including reprinting/republishing this material for advertising or promotional purposes, creating new collective works, for resale or redistribution to servers or lists, or reuse of any copyrighted component of this work in other works. Accepted for publication in the Proceedings of the IEEE International Conference on Data Mining (ICDM) 2026.}}

\begin{document}
\bstctlcite{IEEEexample:BSTcontrol}
\maketitle

\begin{abstract} 
Generative retrieval (GR) for sequential recommendation compresses each item into a short semantic ID and trains an autoregressive decoder to generate next-item codes from a user history. While appealing for cold-start handling and storage, autoregressive GR suffers a structural scaling limit: typically, longer semantic IDs reduce quantization error in principle but
degrade ranking quality and quadratic memory usage against semantic ID length.
We propose \textbf{Late Quantization (LQ)}, a non-autoregressive paradigm that confines discretization to the output (item) side of the pipeline. The encoder and scoring rule operate in continuous space, while items are stored and matched as semantic IDs produced by an end-to-end-trained residual quantizer.
This architectural shift permits an additive error decomposition and a pairwise ordering theorem that jointly bound the quantization-induced ranking drop, a structural guarantee absent in prior semantic-ID recommenders. Across four public benchmarks, the resulting recommender, \textbf{LQ-Rec}, delivers three capabilities simultaneously: ranking quality improves monotonically with the semantic-ID length and outperforms existing baselines; a single trained model supports anytime compression, tuning ranking quality against inference cost without retraining; and LQ-Rec shows superior generalization on a memorization/generalization partition.
\end{abstract}


\section{Introduction}
\label{sec:introduction}

Generative retrieval (GR), originally proposed for document
retrieval~\cite{dsi,nci}, is gathering attention as an emerging
paradigm for sequential recommendation~\cite{tiger,letter,lcrec,rpg}.
Typically, GR assigns each item a short tuple of discrete codes (a \emph{semantic
ID}) and trains an autoregressive decoder to generate the next item
code sequence given a user history.
As argued by Rajput et al.~\cite{tiger}, the semantic ID
representation avoids storing a full embedding per item, and enables
cold-start handling through a content encoder and the hierarchical ID structure of residual quantization.

However, autoregressive GR exhibits a structural
problem: the sequence encoder consumes the pretrained quantized item codes, so
the predictor and the quantizer cannot be analyzed or optimized in
isolation.
This coupling has three concrete consequences.
\emph{Scaling:}
Let $M$ denote the number of code tokens per item and $N$ the
history length: encoding each history item as $M$ tokens forces the
encoder to attend over an $MN$-token sequence, with quadratic attention memory
$O((MN)^2)$. 
\emph{Ranking quality:}
As empirically shown in \cite{rpg} (and our Appendix~\ref{app:tiger-depth}), scaling $M$ also
does not guarantee the improvement of 
ranking quality rather than improves it: a longer semantic ID nominally reduces
the per-item quantization error, but the autoregressive prediction
task itself becomes harder. %
\emph{Inference cost:}
Beam-search decoding adds another $O(bM)$ sequence-encoder forward
passes per query, where $b$ is the beam width.
Autoregressive methods are therefore confined in practice to short
semantic IDs, bounding the code length they can spend on retrieval
quality.

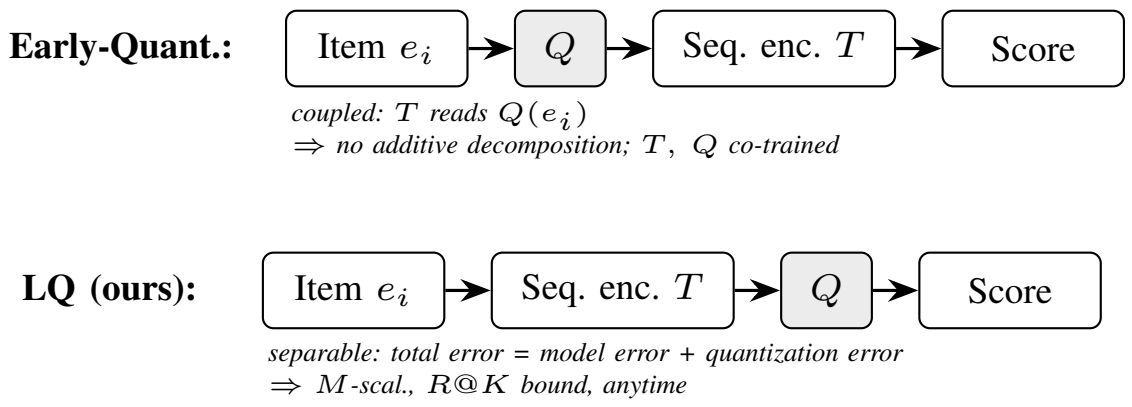
\begin{figure}[t]
\centering
\begin{tikzpicture}[
  font=\footnotesize,
  node distance=3mm,
  box/.style={draw, rounded corners=2pt, minimum height=5mm,
              minimum width=12mm, inner sep=1.5pt, align=center,
              font=\scriptsize},
  qbox/.style={draw, fill=gray!15, rounded corners=2pt,
               minimum height=5mm, minimum width=6mm, inner sep=1pt,
               align=center, font=\scriptsize},
  tbox/.style={draw, rounded corners=2pt, minimum height=5mm,
               minimum width=16mm, inner sep=1.5pt, align=center,
               font=\scriptsize},
  lab/.style={font=\scriptsize\bfseries, inner sep=1pt,
              minimum width=14mm, align=left},
  note/.style={font=\tiny\itshape, inner sep=1pt},
  arrow/.style={-Stealth, semithick},
]
\node[lab] (preqtag) {Early-Quant.:};
\node[box,  right=of preqtag] (pe) {Item $e_i$};
\node[qbox, right=of pe]      (pQ) {$Q$};
\node[tbox, right=of pQ]      (pT) {Seq.\ enc.\ $T$};
\node[box,  right=of pT]      (pS) {Score};
\draw[arrow] (pe) -- (pQ);
\draw[arrow] (pQ) -- (pT);
\draw[arrow] (pT) -- (pS);
\node[note, below=0.5mm of pe.south west, anchor=north west,
       align=left]
  {coupled: $T$ reads $Q(e_i)$\\
   $\Rightarrow$ no additive decomposition; $T,\,Q$ co-trained};
\node[lab, below=16mm of preqtag.west, anchor=west] (lqtag)
  {LQ (ours):};
\node[box,  right=of lqtag] (le) {Item $e_i$};
\node[tbox, right=of le]    (lT) {Seq.\ enc.\ $T$};
\node[qbox, right=of lT]    (lQ) {$Q$};
\node[box,  right=of lQ]    (lS) {Score};
\draw[arrow] (le) -- (lT);
\draw[arrow] (lT) -- (lQ);
\draw[arrow] (lQ) -- (lS);
\node[note, below=0.5mm of le.south west, anchor=north west,
       align=left]
  {separable: total error = model error + quantization error\\
   $\Rightarrow$ $M$-scal., $R@K$ bound, anytime};
\end{tikzpicture}
\caption{\textbf{Position of the quantization step determines
whether the retrieval error is separable.}
Prior semantic-ID methods (top) quantize items
before the sequence encoder, coupling the model and quantization
errors.  LQ (bottom) confines quantization to the
item side, so the total error decomposes additively
(Proposition~\ref{prop:error-decomp})
and admits separable optimization
(Proposition~\ref{prop:separable}).}
\label{fig:lq-vs-preq}
\end{figure}

We propose \textbf{Late Quantization (LQ)}, a non-autoregressive
paradigm that resolves the coupling at its source by reversing the
quantization order: discretization is confined to the \emph{output
side} of the retrieval pipeline, so the encoder and the scoring rule
operate in continuous space and never read discrete tokens
(Fig.~\ref{fig:lq-vs-preq}).
The encoder and a residual item quantizer are nonetheless jointly
trained end-to-end (Section~\ref{sec:method}), so that the continuous
representation is shaped from the outset to be amenable to
discretization.
Because the encoder never reads quantized tokens, the LQ score
decomposes additively into a continuous-model term and a
quantization term (Proposition~\ref{prop:error-decomp}), and the
quantization-induced ranking drop is bounded by per-item
quantization errors and pairwise score margins
(Theorem~\ref{thm:pairwise}).
We call the recommender \textbf{LQ-Rec}.

Like other semantic-ID methods, LQ-Rec stores each item as a short
semantic ID rather than a full continuous embedding, keeping
item-side storage compact at large catalog sizes.
RPG~\cite{rpg} also avoids autoregression and recovers
$M$-scalability by parallel multi-token prediction, but its encoder
still reads quantized inputs, so it does not inherit this additive
decomposition or the bound that follows from it.

We summarize the resulting properties of LQ-Rec as three
capabilities, each a direct consequence of the additive error
decomposition (Propositions~\ref{prop:error-decomp},
\ref{prop:separable}) and each validated on four public benchmarks:

\begin{enumerate}
\item[(i)] \emph{$M$-scalability.}
  Ranking quality improves monotonically with $M$ %
  and dominates published baselines at matched $M$;
  encoder memory is $O(N^2)$ in $M$ rather than $O((MN)^2)$
  (Section~\ref{sec:main}).
\item[(ii)] \emph{Anytime compression} (LQ-unique).
  A single $M$-trained codebook can be truncated to any prefix
  length $L \le M$ at inference time without retraining
  (Section~\ref{sec:anytime}).
\item[(iii)] \emph{Generalization.} LQ-Rec shows superior generalization on the MemGen~\cite{memgen} test partition relative to existing semantic-ID baselines  (Section~\ref{sec:generalization}).
\end{enumerate}

We obtain these capabilities through the following contributions
to the LQ framework:
\begin{itemize}
\item \textbf{Theoretical framework.}
  An additive error decomposition with separable optimization
  (Propositions~\ref{prop:error-decomp},~\ref{prop:separable})
  and a pairwise ordering theorem
  (Theorem~\ref{thm:pairwise}) jointly bound the
  quantization-induced ranking drop --- a structural guarantee
  absent in prior semantic ID-based sequential recommenders.
\item \textbf{Framework and method.}
  LQ confines discretization to the output side while keeping
  prediction in continuous space; we instantiate it as LQ-Rec and
  train it in two stages (S2H-QA + Prefix-MQA) that operationalize
  the separable structure
  (Sections~\ref{sec:lq-framework},~\ref{sec:method}).
\item \textbf{Empirical validation.}
  Our evaluation demonstrates that LQ-Rec matches or exceeds every published baseline on all three capabilities above (Sections~\ref{sec:main}--\ref{sec:generalization}); to our knowledge, this is the first non-autoregressive semantic-ID recommender to deliver all three simultaneously.
\end{itemize}

\section{Related Work}
\label{sec:related}

\emph{ID-based sequential recommenders}~\cite{sasrec,hgn,s3rec,gru4rec,caser,bert4rec}
predict the next item by self-attention or gating over a
per-item embedding lookup; storage scales linearly with the
catalog, and cold or low-frequency items receive underfit
representations.  \emph{Content-based dense retrievers}, such
as UniSRec~\cite{unisrec} and VQ-Rec~\cite{vqrec}, replace
item-ID embeddings with text-encoded content representations to
enable cross-domain transfer and cold-start handling; VQ-Rec
further passes the embedding through optimized product
quantization to obtain a domain-invariant discrete bridge
before dense retrieval.

\emph{Autoregressive generative recommenders (AR-GR)}, introduced
by TIGER~\cite{tiger}, tokenize items as semantic IDs via RQ-VAE~\cite{rqvae}
and generate the next item's tokens autoregressively with a
seq2seq transformer; follow-ups refine the tokenizer
(LETTER~\cite{letter}) or fold the codes into LLM vocabularies
(LC-Rec~\cite{lcrec}).
All inherit AR-GR's compounding decoder error, which caps the
usable semantic-ID length $M$.
RPG~\cite{rpg} sidesteps this with a \emph{non-autoregressive}
parallel multi-token prediction over OPQ-derived codes.  LQ-Rec
also bypasses autoregression and additionally keeps the encoder
in continuous space, yielding the item-wise bound
(Theorem~\ref{thm:pairwise}).

Two orthogonal lines of work inform our evaluation.
fMRLRec~\cite{fmrlrec}, building on Matryoshka Representation
Learning~\cite{mrl}, exposes an anytime accuracy/cost tradeoff by
truncating a continuous item embedding---a different direction of
compression from LQ-Rec's semantic-ID prefix truncation.
Ding et al.~\cite{memgen} introduced a Mem/Gen test split
(training-observed vs.\ unseen transitions); we adopt it in
Section~\ref{sec:generalization}.

Table~\ref{tab:capability-comparison} positions LQ-Rec against the
paradigms above.
Only LQ-Rec admits the additive error decomposition
(Propositions~\ref{prop:error-decomp},~\ref{prop:separable}),
from which both a tractable bound on the quantization-induced
ranking drop (Theorem~\ref{thm:pairwise}) and the anytime
compression at inference (Section~\ref{sec:anytime}) follow.

\begin{table}[t]
\centering
\caption{Capability comparison of sequential recommenders.}
\label{tab:capability-comparison}
\footnotesize
\setlength{\tabcolsep}{3pt}
\begin{tabular}{lcccc}
  \toprule
   & Catalog & $M$- & Additive & Anytime \\
  Method & scal. & scal. & decomp.\ & comp. \\
  \midrule
  ID-based (e.g., SASRec)  & $\times$     & --           & --           & --           \\
  AR-GR (e.g., TIGER)      & $\checkmark$ & $\times$     & $\times$     & $\times$     \\
  RPG                      & $\checkmark$ & $\checkmark$ & $\times$     & $\times$     \\
  \textbf{LQ-Rec (ours)}   & $\checkmark$ & $\checkmark$ & $\checkmark$ & $\checkmark$ \\
  \bottomrule
\end{tabular}
\vspace{-1em}
\end{table}

\section{Setup and Notation}
\label{sec:setup}

\begin{definition}[Recommendation setting]
Let $\mathcal{I} = \{1, \dots, N\}$ be an item catalog.
A \emph{sequential recommendation model} maps a user's interaction history
$(a_1, \dots, a_L) \in \mathcal{I}^L$ to a user vector $u \in \R^d$
via a transformer $T$, and assigns a relevance score to each item
$i$ via
$s(i) \;=\; \inner{u}{e_i}$,
where $e_i \in \R^d$ is the embedding of item $i$.
Items are ranked by decreasing score.
\end{definition}

\begin{definition}[Quantization and quantization error]
A quantizer $Q\colon \R^d \to \R^d$ maps each item embedding to a
finite-precision approximation $\hat{e}_i = Q(e_i)$.
We define:
\begin{itemize}
  \item $\delta_i = \hat{e}_i - e_i$: the quantization error vector for item $i$,
  \item $\varepsilon_i = \norm{\delta_i}$: the quantization error magnitude for item $i$.
\end{itemize}
\end{definition}

\begin{definition}[Residual quantizer]
A residual quantizer of depth $M$ with codebooks
$\{C_1, \dots, C_M\}$, $C_m \in \R^{K \times d}$, encodes an
embedding $e$ as a code tuple $(k_1, \dots, k_M)$ via greedy residual
matching:
\begin{align*}
  r_0 = e, \quad
  r_m = r_{m-1} - C_m[k_m].
\end{align*}
where $k_m = \argmin_{k} \norm{r_{m-1} - C_m[k]}$.
The reconstruction at prefix length $L \le M$ is
$Q_L(e) \;=\; \sum_{m=1}^{L} C_m[k_m]$,
and the prefix quantization error is
$\varepsilon_i^{(L)} = \norm{e_i - Q_L(e_i)}$.
Setting $L = M$ gives the full reconstruction $\hat{e}_i = Q(e_i)$.
\end{definition}

\begin{definition}[Pairwise score difference]
For an ordered pair $(i,j)$ of items and a user vector $u$:
\begin{align}
  s_{ij}      &= \inner{u}{e_i - e_j}              &&\text{(continuous item)}, \\
  \hat{s}_{ij} &= \inner{u}{\hat{e}_i - \hat{e}_j} &&\text{(quantized item)}.
\end{align}
\end{definition}

\begin{definition}[Pairwise violation rate]
For a user vector $u$, the \emph{violation rate} is the fraction of pairs ranked positively by the continuous score whose ordering is reversed by quantization:
\[
  \mathrm{VR}(u)
  \;=\;
  \Prob_{(i,j):\, s_{ij} > 0}\!\bigl(\hat{s}_{ij} \le 0\bigr).
\]
\end{definition}

\section{Late Quantization: Framework, Architecture, and Theoretical Foundation}
\label{sec:foundation}

This section introduces the Late Quantization framework
(Section~\ref{sec:lq-framework}), instantiates it as LQ-Rec
(Section~\ref{sec:architecture}), and states three results
(Sections~\ref{sec:error-decomp}--\ref{sec:rq-decay}) that ground the
three capability claims (Section~\ref{sec:foundation-to-claims}).

\subsection{The Late Quantization (LQ) framework}
\label{sec:lq-framework}

LQ is a deliberately simple three-stage pipeline.
\emph{(i)~Continuous model:} a continuous sequential recommender
provides the upstream representation: a transformer $T$ produces a user vector $u = T(h)$ from history $h$ and items are scored by inner product $s(i) = \inner{u}{e_i}$. %
\emph{(ii)~Quantization:} a quantizer $Q\colon \R^d \to \R^d$ maps each continuous item embedding to a compact, codebook-encoded representation $\hat{e}_i = Q(e_i)$; the resulting index tuple $(k_1, \dots, k_M)$ is the semantic ID of item $i$.
\emph{(iii)~Quantized inference:} the recommender invokes the sequence encoder once to obtain $u = T(h)$ and ranks items by inner product against the quantized table:
$\hat{s}(i) = \inner{u}{\hat{e}_i}$.
Schematically,
\begin{equation}\label{eq:lq-pipeline}
  \underbrace{h \;\xrightarrow{\;T\;}\; u}_{\text{sequence encoder}}
  \qquad
  \underbrace{e_i \;\xrightarrow{\;Q\;}\; \hat{e}_i}_{\text{item quantizer}}
  \qquad
  \underbrace{\hat{s}(i) \;=\; \inner{u}{\hat{e}_i}}_{\text{inner-product scoring}}.
\end{equation}

\noindent
\emph{``Late''} refers to the position of the quantization step
relative to prediction: the model first learns a coherent continuous
representation, and discretization is applied \emph{afterwards}.
This is in deliberate contrast to autoregressive generative retrieval,
where the predictive model is forced to emit discrete code tokens
directly and the prediction task itself becomes harder as the code
length $M$ grows.
We instantiate this framework as LQ-Rec, described next; the
two-stage training pipeline that produces the $(T, Q)$ pair for
LQ-Rec is deferred to Section~\ref{sec:method}.

\subsection{LQ-Rec architecture}
\label{sec:architecture}

\begin{figure}[t]
\centering
\includegraphics[width=0.95\linewidth]{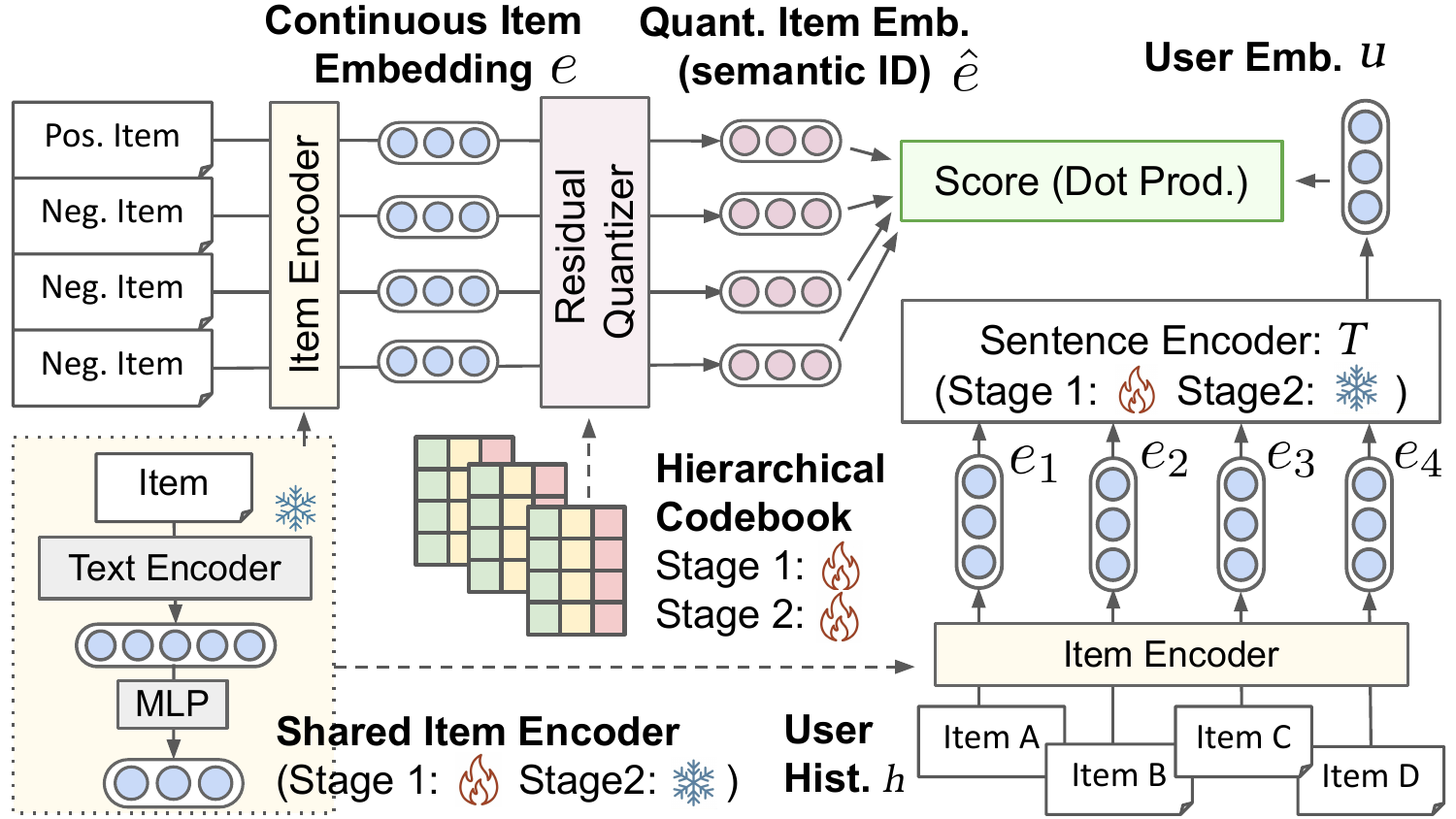}
\caption{LQ-Rec architecture.  The item path
$\text{text\_emb}_i \xrightarrow{\mathrm{MLP}} e_i$ is the only
item-side module: the sequence encoder $T$ reads the user history
$h$ as the sequence of item embeddings $(e_{a_t})_{t=1}^{L}$ produced
by this path and returns the user vector $u$, while the same $e_i$
on the candidate side is encoded by a residual quantizer $Q$ into a
semantic ID $(k_1, \dots, k_M)$ with continuous reconstruction
$\hat{e}_i$.  At inference, items are ranked by
$\hat{s}(i) = \inner{u}{\hat{e}_i}$.}
\label{fig:architecture}
\end{figure}

LQ-Rec instantiates the three components of the LQ pipeline
(Eq.~\eqref{eq:lq-pipeline}) as follows (see Fig.~\ref{fig:architecture} for overall data flow).

\subsubsection{Item encoder}
Each item $i$ is encoded by a frozen text encoder $\text{text\_emb}_i \in \R^{d_{\text{text}}}$;
a learnable text-MLP projects it to the transformer width:
$e_i \;=\; \mathrm{MLP}(\text{text\_emb}_i) \;\in\; \R^H$,
which is the single continuous item embedding used by both training
and inference. No per-item ID embedding is learned, so
LQ-Rec is fully content-based.

\subsubsection{Sequence encoder $T$}
Given a user's interaction history $h = (a_1, \dots, a_L)$,
the corresponding sequence of continuous item embeddings $(e_{a_1}, \dots, e_{a_L})$ is fed into the decode-only transformer blocks, and the final-position hidden state is returned as the user
vector $u \in \R^H$.
Backbone hyperparameters ($H$, depth, heads, dropouts, etc.) are fixed in Section~\ref{sec:setup-exp}.

\subsubsection{Residual item quantizer $Q$}
$Q$ is the residual quantizer of Section~\ref{sec:setup}, with
codebook width matched to the transformer hidden size
($C_m \in \R^{K \times H}$) and applied to the item encoder output
$e_i$.  It produces the semantic ID $(k_1^i, \dots, k_M^i)$ stored
on disk and the continuous reconstruction $\hat{e}_i = Q(e_i)$ used
for scoring.  The residual structure further exposes the family of
prefix quantizers $\{Q_L\}_{L=1}^{M}$ that LQ-Rec exploits for
anytime compression (Section~\ref{sec:method-stage2}).

\subsubsection{Inference}
A single transformer forward pass returns $u = T(h)$.  Items are
then retrieved by beam search over the semantic-ID prefix tree:
because $\hat{e}_i = \sum_{m=1}^{M} C_m[k_m^i]$ decomposes
additively, each beam $(k_1, \dots, k_m)$ carries the partial score
$\sum_{m'=1}^{m} \inner{u}{C_{m'}[k_{m'}]}$, and the depth-$m$
expansion ranks $K$ codebook extensions in a single
$K \times H$ centroid inner-product lookup, with the catalog prefix
trie pruning extensions that have no valid item completion.
Unlike autoregressive GR, beam expansion adds no further transformer
forward passes; the encoder is invoked exactly once per query, and
the depth-$M$ score $\hat{s}(i) = \inner{u}{\hat{e}_i}$ coincides
with the dense inner product against the quantized item table.
Moreover, since the residual decomposition is additive, we obtain
the \textbf{anytime compression} feature (see Section~\ref{sec:foundation-to-claims}).

\subsection{Additive error decomposition}
\label{sec:error-decomp}

In Late Quantization the retrieval error decomposes at the score
level into a \emph{model} term and a \emph{quantization} term that
can be analyzed independently.
Fix a user history $h$ producing user vector $u = T(h)$ and let
$s^\star(i) \in \R$ denote the (unobservable) ground-truth utility of
item $i$ for this user---the quantity that the served score is
ultimately trying to approximate.
The LQ pipeline of Eq.~\eqref{eq:lq-pipeline} produces two
intermediate scores: the continuous model score
$s(i) = \inner{u}{e_i}$ and the quantized model score
$\hat{s}(i) = \inner{u}{\hat{e}_i}$.

\begin{proposition}[Score-level error decomposition]\label{prop:error-decomp}
Under the LQ pipeline of Eq.~\eqref{eq:lq-pipeline}, the quantized
score decomposes linearly into the continuous score and a per-item
quantization perturbation,
\[
  \hat{s}(i) \;=\; s(i) \;+\; \inner{u}{\delta_i},
\]
so the pointwise error against the ground-truth utility $s^\star(i)$
splits additively as
\[
  \underbrace{s^\star(i) - \hat{s}(i)}_{\text{Total Error}}
  \;=\;
  \underbrace{\bigl(s^\star(i) - s(i)\bigr)}_{\mathcal{E}_M(i)}
  \;+\;
  \underbrace{\bigl(-\inner{u}{\delta_i}\bigr)}_{\mathcal{E}_Q(i)}.
\]
\end{proposition}

\begin{remark}[The decomposition is LQ-specific]\label{rem:lq-specific}
The linear score decomposition $\hat s(i) = s(i) + \inner{u}{\delta_i}$
hinges on $u = T(h)$ being invariant to $Q$.
Early Quantization (TIGER, LETTER, VQ-Rec) reads code tokens through
a code-embedding table tied to $Q$, and the scoring rule is not even
an inner product against a continuous $\hat e_i$ (TIGER's is an
autoregressive likelihood over codes).
There is no continuous companion $s(i)$ to which $\hat s(i)$ can be
referred, so no analogue of the additive decomposition or of the
bound $|\mathcal{E}_Q(i)| \le \|u\|\,\|\delta_i\|$ exists, and
Theorem~\ref{thm:pairwise} and Corollary~\ref{cor:recall} are
intrinsically LQ properties.
\end{remark}

\textbf{Model Error} $\mathcal{E}_M(i) = s^\star(i) - s(i)$ is the
gap between the ground-truth utility and the continuous
transformer-plus-MLP score; it is governed by the recommendation
training objective, model capacity, content features, and the data
distribution---i.e., everything the continuous backbone is in charge
of.
\textbf{Quantization Error}
$\mathcal{E}_Q(i) = -\inner{u}{\delta_i}$ is what quantization adds
on top of the continuous score; it is governed by the codebook
(through $\delta_i$) and the code length $L$ at inference.

Theorem~\ref{thm:pairwise} lifts this score-level
decomposition to a per-pair ranking bound that translates a
controlled per-item quantization error $\varepsilon_i$ into a
controlled Recall@$K$ drop.

\begin{theorem}[Pairwise Ordering Preservation]\label{thm:pairwise}
Let $u \ne 0$ be the user vector and $(i,j)$ a pair with
$s_{ij} > 0$.
Write $\udir = u/\norm{u}$ and define the \emph{normalized margin}
$m_{ij} = s_{ij}/\norm{u} = \inner{\udir}{e_i - e_j}$.
Then, we have the following:
\begin{enumerate}
\item[\textup{(a)}]
  The ordering is preserved ($\hat{s}_{ij} > 0$) whenever
  \(
    m_{ij}
    >
    \varepsilon_i + \varepsilon_j.
  \)
  This condition is independent of $\norm{u}$.
\item[\textup{(b)}]
  The violation rate satisfies
  \(
    \mathrm{VR}(u)
    \le
    \Prob_{(i,j):\, s_{ij}>0}\!\bigl(m_{ij} < \varepsilon_i + \varepsilon_j\bigr).
  \)
\end{enumerate}
\end{theorem}

\begin{proof}[Proof sketch]
Triangle inequality and Cauchy--Schwarz give
$|\hat{s}_{ij} - s_{ij}| \le \norm{u}(\varepsilon_i + \varepsilon_j)$;
dividing by $\norm{u}$ yields~(a), and~(b) follows.
Full proof in Appendix~\ref{app:proofs}.
\end{proof}

The theorem states that the score ordering is preserved as long as
the quantization errors are small and the score margins are large.
Writing $\mathrm{Recall@}K(\mathrm{cont})$ and
$\mathrm{Recall@}K(\mathrm{LQ})$ for the population Recall@$K$
computed with the continuous scores $\{s(i)\}$ and the quantized
scores $\{\hat{s}(i)\}$ respectively, the following corollary
translates this pairwise statement into a Recall@$K$ bound (proof in Appendix~\ref{app:proofs}).
We discuss the empirical tightness of this worst-case bound in Appendix~\ref{app:rankloss-tightness}.

\begin{corollary}[Recall@$K$ bound]\label{cor:recall}
Let $i^*$ be in the continuous top-$K$ for user $u$.
Define the \emph{error-adjusted minimum margin}
\[
  \tilde{\gamma}_{i^*}
  \;=\;
  \min_{j:\,\mathrm{rank}_u(j) > K}
  \bigl(m_{i^* j} - \varepsilon_{i^*} - \varepsilon_j\bigr).
\]
Then $i^*$ remains in the quantized top-$K$ whenever
$\tilde{\gamma}_{i^*} > 0$.
Averaging over the population:
\[
  \mathrm{Recall@}K(\mathrm{LQ})
  \;\ge\;
  \mathrm{Recall@}K(\mathrm{cont})
  \;-\;
  \Prob_{(u,\,i^*)}\!\bigl(\tilde{\gamma}_{i^*} \le 0\bigr).
\]
\end{corollary}

The corollary gives two levers for limiting the
quantization-induced Recall@$K$ degradation: shrinking the
per-item quantization errors $\varepsilon_{i^*}, \varepsilon_j$,
and enlarging the score margins $m_{i^* j}$.

\subsection{Separable training under LQ}
\label{sec:separable}

Because the sequence encoder never reads quantized tokens, the LQ
pipeline also makes the training objective \emph{separable}: the
encoder parameters can be optimized under a continuous loss, and the
codebook parameters under a quantized loss, as distinct sub-problems.

\begin{proposition}[Separable training under LQ]\label{prop:separable}
Let $\Theta_T$ denote the parameters of $T$ and the item MLP, and
$\Theta_Q$ the codebook parameters of $Q$.
Under the LQ pipeline of Eq.~\eqref{eq:lq-pipeline}, the user vector
$u = T(h)$ and the continuous score $s(i) = \inner{u}{e_i}$ are
functions of the continuous embeddings $\{e_i\}$ only and do not
depend on $\Theta_Q$.
Consequently:
\begin{enumerate}
\item[\textup{(a)}] any continuous-side loss
  $\mathcal{L}_{\mathrm{cont}} = \ell(s^\star, s)$ is
  $\Theta_Q$-independent and updates only $\Theta_T$;
\item[\textup{(b)}] any quantized-side loss
  $\mathcal{L}_{\mathrm{quant}} = \ell(s^\star, \hat{s})$ evaluated
  with $\Theta_T$ held fixed reduces to a function of $\Theta_Q$
  alone, since $u = T(h)$ is computable without reference to $Q$.
\end{enumerate}
Training procedures that update $\Theta_T$ under a continuous loss
and $\Theta_Q$ under a quantized loss with $\Theta_T$ frozen are
therefore well-defined separable sub-problems.
\end{proposition}

\begin{proof}
By Eq.~\eqref{eq:lq-pipeline}, $T$ reads the continuous sequence
$(e_{a_t})_{t=1}^L$ rather than the quantized
$(\hat{e}_{a_t})_{t=1}^L$, so no factor of $Q$ enters $u$ or $s(i)$;
both (a) and (b) follow immediately by substitution into
Proposition~\ref{prop:error-decomp}.
\end{proof}

\begin{remark}[EQ is not separable]\label{rem:preq-coupled}
As noted in Remark~\ref{rem:lq-specific}, the EQ encoder reads
code tokens, so the scoring rule depends on $\Theta_Q$ end-to-end;
$\mathcal{L}_{\mathrm{cont}}$ is therefore not $\Theta_Q$-independent
and the codebook and predictor must be co-trained under a single
coupled objective.
\end{remark}

\begin{remark}[Dependence on inner-product scoring]
The theoretical results above assume an inner-product scoring rule, $s(i) = \langle u, e_i \rangle$. While this is standard and highly efficient for retrieval tasks, extending the LQ framework to non-linear scoring functions (e.g., deep interaction MLPs) breaks this exact additive separability. We leave the adaptation of LQ to such complex ranking models as future work.
\end{remark}

\subsection{Theoretical implications}
\label{sec:foundation-to-claims}

Proposition~\ref{prop:error-decomp} and
Theorem~\ref{thm:pairwise} underpin the three capability claims
of Section~\ref{sec:introduction} as follows.

\noindent\textbf{(i)~$M$-scalability $\leftrightarrow$ Proposition~\ref{prop:error-decomp}.}
GR couples quantization and prediction; LQ does not.
The autoregressive decoder of TIGER predicts a length-$M$ code
sequence step by step, so the prediction task itself becomes
harder with $M$~\cite{rpg}.
In LQ, Proposition~\ref{prop:error-decomp} isolates the only
$M$-dependent term in the score: $s(i) = \inner{u}{e_i}$ is
$M$-independent (the prediction task is a fixed continuous inner
product on an $H$-dimensional vector), and the entire $M$ dependence
is confined to $\inner{u}{\delta_i}$.
Under any reasonable training procedure, the quantization error
$\varepsilon_i = \norm{\delta_i}$ is non-increasing in $M$, so retrieval
quality is non-decreasing in $M$ for the same continuous predictor
(Corollary~\ref{cor:recall}).

\noindent\textbf{(ii)~Anytime compression $\leftrightarrow$ Proposition~\ref{prop:error-decomp}.}\label{sec:rq-decay}
The residual structure of $Q$ exposes a family of quantizers
$Q_L$ parameterized by prefix length $L \le M$, with $Q_M = Q$.
The same argument applied to each $Q_L$ --- with $L$ in place of $M$
--- gives retrieval quality non-decreasing in $L$ for a single
trained codebook.
The remaining question is whether that codebook can be predictive
at every $L$ simultaneously; Section~\ref{sec:method-stage2} answers
this via the prefix-MRL Stage~2 objective, and
Section~\ref{sec:anytime} provides the empirical curve.

\noindent\textbf{(iii)~Generalization $\leftrightarrow$ Theorem~\ref{thm:pairwise}.}
Because Theorem~\ref{thm:pairwise} is a per-pair statement, it
applies unchanged on the test split. 
Together with the additive
decomposition (Proposition~\ref{prop:error-decomp}), this
disentangles two design responsibilities: the continuous backbone
determines Recall@K(cont), and the residual quantizer determines
the quantization-induced gap. %
Any architectural property held by the continuous backbone---in particular,
content-based generalization to cold and low-frequency items, and
to the test transitions of the MemGen
partition~\cite{memgen}---is inherited by the quantized model
modulo a bounded gap.
Section~\ref{sec:generalization} verifies this empirically.

\section{Two-Stage Training}
\label{sec:method}

LQ-Rec is trained in two stages.
\textbf{Stage~1} (\emph{S2H-QA}, Soft-to-Hard Quantization-Aware training)
jointly trains $T$ and $Q$ end-to-end under a scheduled convex
combination of a continuous (soft) and a quantized (hard) ranking
loss, annealing from soft to hard.
\textbf{Stage~2} (\emph{Prefix-MQA}, Prefix-Matryoshka Quantization-Aware
fine-tuning) freezes $T$ and fine-tunes the codebook, realizing the
separable sub-problem of Proposition~\ref{prop:separable}(b), with
cross-entropy losses computed at every prefix length
$L = 1, \dots, M$ (truncation along the RQ depth axis, not the
embedding-dimension axis as in standard MRL~\cite{mrl}).

\subsection{Stage~1 Training: S2H-QA}
\label{sec:method-stage1}

For each training step, we run \emph{one} transformer forward pass
to produce the user vector $u$.
On the item side, the architecture of Section~\ref{sec:architecture}
provides the continuous embedding
$e_i = \mathrm{MLP}(\text{text\_emb}_i)$;
Stage~1 additionally forms a straight-through (STE) quantized view of
the same embedding,
\[
  \hat{e}_i \;=\; e_i + \bigl(q_i - \sg(e_i)\bigr),
  \qquad q_i = Q(e_i),
\]
which is equal to $q_i$ in the forward pass and yields gradient
$\partial \hat{e}_i / \partial e_i = 1$ and
$\partial \hat{e}_i / \partial q_i = 1$ in the backward pass.
Two score vectors and two cross-entropy losses are formed:
\begin{align*}
  \hat{s}_e(i) &= \inner{u}{e_i},
  & \mathcal{L}_{\mathrm{cont}}  &= \mathcal{L}_{\mathrm{CE}}(\hat{s}_e), \\
  \hat{s}_q(i) &= \inner{u}{\hat{e}_i},
  & \mathcal{L}_{\mathrm{quant}} &= \mathcal{L}_{\mathrm{CE}}(\hat{s}_q),
\end{align*}
both evaluated against the same positive label $i^*$ and the same user
vector $u$.

The Stage~1 objective is the convex combination
\[
  \mathcal{L}_{\mathrm{mix}}(\alpha)
  \;=\;
  (1 - \alpha)\,\mathcal{L}_{\mathrm{cont}}
  \;+\;
  \alpha\,\mathcal{L}_{\mathrm{quant}}
  \;+\;
  \alpha\,\beta\,\underbrace{\norm{e_i - \sg(q_i)}^2}_{\mathcal{L}_{\mathrm{commit}}},
\]
where $\alpha(t) \in [0, 1]$ is a scheduled mixing coefficient and
$\beta \ge 0$ controls the commitment regularizer.
At $\alpha = 0$ the objective reduces to standard CE on continuous
embeddings (a vanilla SASRec step); at $\alpha = 1$ it is CE on the
quantized embeddings, exactly matching the inference-time forward
pass.
Whenever $\alpha < 1$, $\mathcal{L}_{\mathrm{cont}}$ contributes a
continuous-side gradient on the encoder that is independent of the
codebook state (Proposition~\ref{prop:separable}(a)), letting the
encoder converge toward a coherent continuous geometry before being
asked to operate entirely under STE.

\subsubsection{Schedule}

Training proceeds in three phases:

\begin{itemize}
\item \textbf{Stage~1-A} (continuous warmup, $\alpha = 0$):
  $t \in [0, T_A)$.
  The codebook is uninitialized; only $\mathcal{L}_{\mathrm{cont}}$
  is active and the objective collapses to standard SASRec CE.
\item \textbf{$k$-means transition} (single step at $t = T_A$):
  the encoder is snapshotted, and the RQ codebooks are initialized by
  $k$-means on the per-stage residuals of the snapshot embeddings.
\item \textbf{Stage~1-B ramp} ($t \in [T_A, T_P)$):
  $\alpha(t)$ follows a cosine ramp $0 \to 1$.
  The encoder gradually transfers capacity from
  $\mathcal{L}_{\mathrm{cont}}$ to $\mathcal{L}_{\mathrm{quant}}$.
\item \textbf{Stage~1-B plateau} ($t \in [T_P, T_{\max})$):
  $\alpha = 1$.
  $\mathcal{L}_{\mathrm{cont}}$ is zeroed out, and
  $\mathcal{L}_{\mathrm{quant}} + \beta \mathcal{L}_{\mathrm{commit}}$
  is the sole signal.
\end{itemize}
Evaluation for early stopping always runs with $\alpha$ pinned to $1$, so every eval
during Stage~1-B measures the quantized-inference forward pass
regardless of where the training-time $\alpha$ schedule currently
sits. Best-checkpoint selection considers all of Stage~1-B (ramp
\emph{and} plateau); we do not evaluate the model during Stage~1-A evaluations, where the codebook is not
yet initialized and the quantized forward pass is undefined.

\subsubsection{Role in the Pairwise Ordering framework}

The three components of $\mathcal{L}_{\mathrm{mix}}$ act on the
sufficient condition of Theorem~\ref{thm:pairwise} through distinct
mechanisms: $\mathcal{L}_{\mathrm{cont}}$ enlarges the margin
$m_{ij}$ (LHS), $\mathcal{L}_{\mathrm{commit}}$ shrinks the per-item
errors $\varepsilon_i$ (RHS), and $\mathcal{L}_{\mathrm{quant}}$
directly enforces $\hat{s}_{ij} > 0$.
The first two operate on the bound; the third bypasses it by
optimizing the exact directional perturbation
$\inner{u}{\delta_i - \delta_j}$ rather than its Cauchy--Schwarz
upper envelope.
The full bound-bypass analysis is deferred to
Appendix~\ref{app:rankloss-tightness}; we refer back to this routing
in the results sections to explain ablation outcomes.

\subsection{Stage~2 Training: Prefix-MQA}
\label{sec:method-stage2}

Stage~2 takes the Stage~1 encoder and codebook as inputs, freezes the
encoder (and the output projection and temperature), and fine-tunes
the codebook with the prefix-Matryoshka cross-entropy objective alone:
no MSE pull-back $\norm{e_i - \hat{e}_i}^2$ for codebook parameters is added, since that would
constrain the codebook away from task-optimal placement
(Appendix~\ref{app:rankloss-taskopt}).

\subsubsection{Prefix-MRL cross-entropy}

For each prefix length $L \in \{1, \dots, M\}$, define the truncated
quantized embedding
\(
  \hat{e}_i^{(L)} = Q_L(e_i) = \sum_{m=1}^{L} C_m[k_m^i]
\)
and the corresponding score $\hat{s}_q^{(L)}(i) = \inner{u}{\hat{e}_i^{(L)}}$.
The prefix-MRL CE loss is
\begin{equation}\label{eq:prefix-mqa}
  \mathcal{L}_{\mathrm{MRL}\text{-}\mathrm{QA}}
  \;=\;
  \frac{1}{M}\,\sum_{L=1}^{M}
    \mathcal{L}_{\mathrm{CE}}\!\bigl(\hat{s}_q^{(L)}; \tau_L\bigr),
\end{equation}
where $\tau_L = \tau_{\mathrm{base}} \cdot \exp(\Delta_L)$ is a per-$L$
temperature with $\Delta_L$ a trainable scalar (initialized at $0$,
$M$ extra parameters in total).

The per-$L$ temperature $\tau_L$ is introduced because at small $L$
the embedding norm $\norm{\hat{e}_i^{(L)}}$ is reduced, shrinking
logit scales and flattening the softmax; the trainable $\Delta_L$
gives the optimizer a per-branch calibration knob.
Summing over all $L$ in a single objective drives one codebook to be predictive at every prefix length simultaneously, enabling anytime compression. %

\subsubsection{Carry-over codebook initialization}

Stage~2 initializes the codebook by \emph{carry-over} from Stage~1
(re-using the Stage~1 codebook as-is), rather than re-running
$k$-means on the frozen encoder's embeddings.
Carry-over preserves the ranking-specific structure that Stage~1's
$\mathcal{L}_{\mathrm{quant}}$ has already encoded into the codebook;
re-running $k$-means would replace this with the reconstruction-MSE
basin, which is empirically a worse starting point for ranking
(Section~\ref{sec:setup-exp}).

\section{Experiments}
\label{sec:experiments}

We answer the following six research questions:\\
\textbf{RQ1}: Does LQ-Rec outperform existing models?\\
\textbf{RQ2}: How does LQ-Rec performance vary across multiple code lengths?\\
\textbf{RQ3}: Does LQ-Rec show superior generalization on the memorization/generalization partition?\\
\textbf{RQ4}: How does each component of the two-stage training
contribute?\\
\textbf{RQ5}: Is LQ-Rec efficient and scalable at inference? \\ %
\textbf{RQ6}: What is the underlying structure of the learned LQ-Rec semantic IDs?

\subsection{Setup}
\label{sec:setup-exp}

\subsubsection{Datasets}
We use three Amazon Review categories (Toys, Beauty, Sports) with leave-one-out
evaluation (5-core), following Rajput et al.~\cite{tiger}. Item content is the
concatenation of title, brand, category, and description. We additionally include
Yelp 5-core, where the item content is the title plus categories.
All four datasets use the preprocessed 5-core release distributed in  \url{https://github.com/jeykigung/P5}~\cite{p5}; per-dataset statistics are summarized in Table~\ref{tab:dataset-stats}.

\begin{table}[t]
\centering
\caption{Dataset statistics after 5-core preprocessing.}
\label{tab:dataset-stats}
\footnotesize
\setlength{\tabcolsep}{6pt}
\begin{tabular}{l r r r r}
\toprule
Dataset & \#users & \#items & \#interactions & avg.\ seq.\ len.\ \\
\midrule
Toys   & 19{,}412 & 11{,}924 & 167{,}597 & 8.63 \\
Beauty & 22{,}363 & 12{,}101 & 198{,}502 & 8.88 \\
Sports & 35{,}598 & 18{,}357 & 296{,}337 & 8.32 \\
Yelp   & 30{,}431 & 20{,}033 & 316{,}354 & 10.40 \\
\bottomrule
\end{tabular}
\vspace{-0.5em}
\end{table}

\subsubsection{Baselines}
We compare LQ-Rec with the following non semantic ID-based and semantic ID-based baselines:\\
  \textbf{HGN}~\cite{hgn}: hierarchical feature- and instance-level gating over item embeddings.
  \textbf{SASRec}~\cite{sasrec}: transformer decoder with self-attention and causal masking.
  \textbf{S$^3$-Rec}~\cite{s3rec}: self-supervised pre-training via mutual information maximization across attribute/item/segment/sequence views, fine-tuned as SASRec.
  \textbf{UniSRec}~\cite{unisrec}: feature-based method with parametric whitening and a MoE adaptor over frozen BERT.
  \textbf{TIGER}~\cite{tiger}: generative retrieval with RQ-VAE semantic IDs and a T5-style seq2seq decoder.
  \textbf{LC-Rec}~\cite{lcrec}: collision-free RQ-VAE (Sinkhorn-Knopp uniform last level) folded into the LLaMA vocabulary via alignment tuning.
  \textbf{LETTER}~\cite{letter}: RQ-VAE tokenizer with CF, diversity, and ranking-aware regularizers.
  \textbf{VQ-Rec}~\cite{vqrec}: dense retrieval with OPQ codes over frozen text encoder.
  \textbf{RPG}~\cite{rpg}: non-autoregressive GR with OPQ-derived independent codes, parallel multi-token prediction, and graph-based iterative decoding.
We mainly used the official implementations or RecBole~\cite{recbole} for baseline implementation, used Ding et al.~\cite{memgen} for TIGER (the best impl. publicly available), and the other baseline results from Hou et al.~\cite{rpg}.

\subsubsection{Metrics}
We report Recall@$K$ and NDCG@$K$ at $K \in \{5, 10\}$ on the
leave-one-out test split.
We train LQ-Rec with 5 seeds and report the mean$\pm$stdev across seeds.

\subsubsection{Implementation details}
For the transformer, we used $H=128$, $L=2$ blocks, $A=2$ attention heads, FFN width $128$, attention dropout $0.2$, output dropout $0.1$, MLP dropout $0.3$, and Pre-RMSNorm.
We used BGE-base-en-v1.5 for the text encoder (768 dim), with the item encoder MLP (768 $\to$ 768 $\to$ 128, GELU activation, dropout=0.3).
The text item embedding is scaled to match the transformer's working width: $\tilde{x} = \sqrt{H}\,\cdot\, x / \|x\|$
($\sqrt{H} \approx 11.31$ at $H = 128$).
We used AdamW
($\mathrm{lr} = 10^{-3}$, weight decay $0.01$, warmup ratio $0.05$),
batch size $256$, evaluating every $100$ steps and early-stopping with patience $20$
evaluations on Recall@10.

We tune two Stage~1 knobs on a per-dataset basis: the commitment-loss
weight $\beta \in \{0.25, 0.5, 1.0\}$ and the codebook learning-rate
multiplier $\mathrm{mult} \in \{1, 2, 5\}$ (codebook LR
$= \mathrm{mult} \cdot 10^{-3}$).  All other Stage~1 settings are fixed:
$M = 16$, $K = 256$, FAISS k-means initialisation, $T_{\max} = 20{,}000$
steps, AdamW with $\text{lr}{=}10^{-3}$, batch size~$256$, a frozen temperature $\tau = 0.5$,
evaluating every $100$ steps and early-stopping with patience $20$ evaluations on Recall@10.
For annealing $\alpha$, we used $T_A = 5{,}000$ and $T_P = 10{,}000$ steps.
Stage~2 is held fixed at
the chain Prefix-MQA recipe ($\text{lr}{=}5\cdot 10^{-4}$, $5{,}000$
steps, encoder frozen).
We used a single NVIDIA 5090 GPU; LQ-Rec training took 18~min (Stage~1) and 19~min (Stage~2).

\subsection{Main results (RQ1)}
\label{sec:main}

\begin{table*}[!t]
\centering
\caption{Comparison of Ranking Metrics (R@$K$~=~Recall@$K$, N@$K$~=~NDCG@$K$).}
\label{tab:main-results}
\footnotesize
\setlength{\tabcolsep}{2.5pt}
\resizebox{\textwidth}{!}{%
\begin{tabular}{l *{4}{cccc}}
  \toprule
  & \multicolumn{4}{c}{Toys} & \multicolumn{4}{c}{Beauty}
  & \multicolumn{4}{c}{Sports} & \multicolumn{4}{c}{Yelp} \\
  \cmidrule(lr){2-5} \cmidrule(lr){6-9} \cmidrule(lr){10-13} \cmidrule(lr){14-17}
  Method
    & R@5 & N@5 & R@10 & N@10
    & R@5 & N@5 & R@10 & N@10
    & R@5 & N@5 & R@10 & N@10
    & R@5 & N@5 & R@10 & N@10 \\
  \midrule
  \multicolumn{17}{c}{\textit{Non-semantic ID-based}} \\
  \midrule
  HGN~\cite{hgn}
    & 0.0321 & 0.0221 & 0.0497 & 0.0277
    & 0.0325 & 0.0249 & 0.0512 & 0.0266
    & 0.0189 & 0.0120 & 0.0313 & 0.0159
    & 0.0132 & 0.0084 & 0.0221 & 0.0111 \\
  SASRec~\cite{sasrec}
    & 0.0463 & 0.0306 & 0.0675 & 0.0374
    & 0.0387 & 0.0249 & 0.0605 & 0.0318
    & 0.0233 & 0.0154 & 0.0350 & 0.0192
    & 0.0153 & 0.0126 & 0.0246 & 0.0126 \\
  S$^3$-Rec~\cite{s3rec}
    & 0.0443 & 0.0294 & 0.0700 & 0.0376
    & 0.0387 & 0.0244 & 0.0647 & 0.0327
    & 0.0251 & 0.0161 & 0.0385 & 0.0204
    & 0.0245 & 0.0206 & 0.0408 & 0.0206 \\
  UniSRec~\cite{unisrec}
    & 0.0477 & 0.0261 & 0.0797 & 0.0364
    & 0.0466 & 0.0269 & 0.0756 & 0.0362
    & 0.0251 & 0.0157 & 0.0426 & 0.0213
    & 0.0279 & 0.0194 & 0.0415 & 0.0238 \\
  \midrule
  \multicolumn{17}{c}{\textit{Semantic ID-based}} \\
  \midrule
  TIGER~\cite{tiger}
    & 0.0438 & 0.0282 & 0.0685 & 0.0361
    & 0.0473 & 0.0313 & 0.0731 & 0.0395
    & 0.0273 & 0.0178 & 0.0432 & 0.0229
    & 0.0240 & 0.0155 & 0.0396 & 0.0205 \\
  LC-Rec~\cite{lcrec}
    & 0.0435 & 0.0294 & 0.0638 & 0.0385
    & 0.0467 & 0.0329 & 0.0685 & 0.0394
    & 0.0318 & 0.0209 & 0.0471 & 0.0301
    & 0.0241 & 0.0158 & 0.0381 & 0.0201 \\
  LETTER~\cite{letter}
    & 0.0447 & 0.0301 & 0.0652 & 0.0398
    & 0.0501 & 0.0355 & 0.0703 & 0.0418
    & 0.0338 & 0.0221 & 0.0489 & 0.0315
    & 0.0255 & 0.0168 & 0.0393 & 0.0211 \\
  VQ-Rec~\cite{vqrec}
    & 0.0555 & 0.0301 & 0.0839 & 0.0393
    & 0.0514 & 0.0294 & 0.0779 & 0.0379
    & 0.0295 & 0.0160 & 0.0466 & 0.0215
    & 0.0373 & \textbf{0.0289} & 0.0512 & 0.0334 \\
  RPG~\cite{rpg}
    & 0.0583 & 0.0395 & 0.0869 & 0.0490
    & 0.0551 & 0.0381 & 0.0809 & 0.0464
    & 0.0295 & 0.0205 & 0.0463 & 0.0263
    & 0.0259 & 0.0176 & 0.0408 & 0.0224 \\
  \midrule
  LQ-Rec
    & \mns{\textbf{0.0653}}{0.0010} & \mns{\textbf{0.0413}}{0.0012} & \mns{\textbf{0.1037}}{0.0013} & \mns{\textbf{0.0537}}{0.0007}
    & \mns{\textbf{0.0640}}{0.0010} & \mns{\textbf{0.0421}}{0.0006} & \mns{\textbf{0.0995}}{0.0007} & \mns{\textbf{0.0535}}{0.0005}
    & \mns{\textbf{0.0396}}{0.0006} & \mns{\textbf{0.0256}}{0.0004} & \mns{\textbf{0.0613}}{0.0005} & \mns{\textbf{0.0323}}{0.0002}
    & \mns{\textbf{0.0400}}{0.0005} & \mns{0.0263}{0.0002} & \mns{\textbf{0.0640}}{0.0004} & \mns{\textbf{0.0340}}{0.0002} \\
  \bottomrule
\end{tabular}%
}
\end{table*}

We compare the two-stage LQ-Rec ($M = L = 16$) against every comparator on the four datasets.
Table~\ref{tab:main-results} shows LQ-Rec at $M = 16$ outperforms all other methods in most cases, with the only exception being VQ-Rec's NDCG@5 on Yelp.
Against RPG, LQ-Rec exceeds Recall@10 by $+19\%$ (Toys),
$+31\%$ (Sports), and $+57\%$ (Yelp) at matched $M = 16$,
and by $+23\%$ on Beauty \emph{at half the $M$}
(RPG uses $M = 32$ there).

\subsection{Anytime compression (RQ2)}
\label{sec:anytime}

A single LQ-Rec ckeckpoint serves arbitrary code lengths at
inference time by prefix truncation $L \in \{1, \dots, M\}$, with
no per-$L$ retraining (Table~\ref{tab:anytime}).
At each prefix length $L$, the same checkpoint is directly comparable to
a separately trained per-$M$ GR baseline at $M = L$.

\begin{table}[t]
\centering
\caption{R@10 at Varying Inference-Time Code Length
  ($K{=}256$).}
\label{tab:anytime}
\footnotesize
\setlength{\tabcolsep}{4pt}
\begin{tabular}{l c c c c}
  \toprule
  Method & Toys & Beauty & Sports & Yelp \\
  \midrule
  \multicolumn{5}{@{}l}{\emph{Per-$M$ baselines (separately trained):}} \\
  \quad TIGER ($M{=}4$)    & 0.0685 & 0.0731 & 0.0432 & 0.0396 \\
  \quad RPG$^{*}$          & 0.0869 & 0.0809 & 0.0463 & 0.0408 \\
  \quad VQ-Rec ($M{=}32$)  & 0.0839 & 0.0779 & 0.0466 & 0.0512 \\
  \midrule
  \multicolumn{5}{@{}l}{\emph{LQ-Rec (single $M{=}16$ ckpt, prefix-truncated):}} \\
  \quad $L = 4$  & 0.0860 & 0.0874 & 0.0511 & 0.0538 \\
  \quad $L = 8$  & 0.0966 & 0.0972 & 0.0585 & 0.0611 \\
  \quad $L = 16$ & 0.1037 & 0.0995 & 0.0613 & 0.0640 \\
  \bottomrule
\end{tabular}
\\[3pt]
{\scriptsize $^{*}$RPG uses per-dataset best $M = 16, 32, 16, 128$ on Toys/Beauty/Sports/Yelp.} 
\vspace{-1em}
\end{table}

LQ-Rec Recall@10 is \emph{monotonically non-decreasing in} $L$ on every
dataset, validating the residual-quantizer monotonicity claim of
Section~\ref{sec:rq-decay}.
At every (baseline, dataset) cell, the same single checkpoint beats the
matching baseline: TIGER ($M = 4$) at $L = 4$ on all four datasets;
RPG at $L = 16$ (at \emph{half} the $M$ on Beauty and at $1/8$ on
Yelp); and VQ-Rec ($M = 32$) at $L = 16$ (\emph{half} the $M$).

\subsection{Generalization (RQ3)}
\label{sec:generalization}

This subsection validates capability~(iii): LQ-Rec shows superior generalization on the MemGen partition relative, and we additionally report cold item results for completeness.

\subsubsection{Mem/Gen partition}
\label{sec:gen-memgen}

The Mem/Gen partition~\cite{memgen} splits each test case
$(u, i^\star)$ by whether $i^\star$ is reachable from the user's
last interaction via a transition observed in training
(\emph{Mem}) or requires structural reasoning such as
substitutability or symmetry (\emph{Gen}).
We compare TIGER ($M = 4$), RPG ($M = 16$/$32$/$16$ on
Toys/Beauty/Sports per the paper recipe), VQ-Rec ($M = 32$), and LQ-Rec
($M = 16$) at prefix lengths $L \in \{4, 8, 16\}$;
each baseline is run at its paper-default $M$.

Table~\ref{tab:gen-memgen} reports per-bucket NDCG@10.
Ding et al.~\cite{memgen} concluded that semantic-ID methods
sacrifice Mem accuracy for stronger Gen relative to ID-based
baselines.  Our results refine this picture: on Gen, LQ-Rec
dominates every dataset, TIGER is next, and the two
ranking-decoupled baselines (RPG and VQ-Rec) are last --- with
VQ-Rec the \emph{worst} Gen retriever, despite sharing LQ-Rec's
content-aware text-encoder family.
The differentiator is therefore neither content-awareness (VQ-Rec is
content-aware too) nor autoregressive decoding (TIGER decodes, RPG
does not, VQ-Rec is dense), but whether the codebook is trained
\emph{jointly with the ranking objective}: VQ-Rec's OPQ centroids
are fixed by reconstruction before the ranking head sees them,
whereas LQ-Rec optimises the codebook end-to-end against the
next-item loss (capability~(i)).
Both LQ-Rec buckets rise monotonically with $L$, so the Gen-side
lead is not specific to any single prefix length.

\begin{table}[!t]
\centering
\footnotesize
\caption{The Mem/Gen Partition Result (N@10)}
\label{tab:gen-memgen}
\setlength{\tabcolsep}{2pt}
\begin{tabular}{l cc cc cc cc}
\toprule
& \multicolumn{2}{c}{Toys}
& \multicolumn{2}{c}{Beauty}
& \multicolumn{2}{c}{Sports}
& \multicolumn{2}{c}{Yelp} \\
\cmidrule(lr){2-3}\cmidrule(lr){4-5}\cmidrule(lr){6-7}\cmidrule(lr){8-9}
& Mem & Gen & Mem & Gen & Mem & Gen & Mem & Gen \\
\midrule
TIGER             & 0.2346 & 0.0245 & 0.2654 & 0.0224 & 0.1546 & 0.0171
                    & 0.1083 & 0.0161 \\
VQ-Rec            & 0.3480 & 0.0156 & 0.3307 & 0.0128 & 0.2538 & 0.0093
                    & \textbf{0.2457} & 0.0223 \\
RPG               & \textbf{0.3888} & 0.0215 & \textbf{0.3595} & 0.0179 & \textbf{0.2578} & 0.0137
                    & 0.1551 & 0.0163 \\
\midrule
LQ-Rec$_{4}$      & 0.3074 & 0.0292 & 0.2804 & 0.0261 & 0.1918 & 0.0197
                    & 0.1633 & 0.0219 \\
LQ-Rec$_{8}$      & 0.3562 & 0.0311 & 0.3300 & 0.0285 & 0.2284 & 0.0229
                    & 0.1980 & 0.0256 \\
LQ-Rec$_{16}$     & 0.3709 & \textbf{0.0313}
                    & 0.3510 & \textbf{0.0293}
                    & 0.2314 & \textbf{0.0240}
                    & 0.2108 & \textbf{0.0267} \\
\bottomrule
\end{tabular}
\vspace{-1em}
\end{table}

\subsubsection{Frequency-stratified retention}
\label{sec:gen-freq}

We stratify each test case by the training-set frequency of the target
item $i^\star$ into four buckets: $\le 5$, $6$--$10$, $11$--$50$, and
$51+$. The same six methods as Table~\ref{tab:gen-memgen} are
evaluated on the same checkpoints.
This stratification cuts at the \emph{item} level and is complementary to the Mem/Gen split, which cuts at the \emph{transition} level.

\begin{figure}[!t]
\centering
\includegraphics[width=0.9\linewidth]{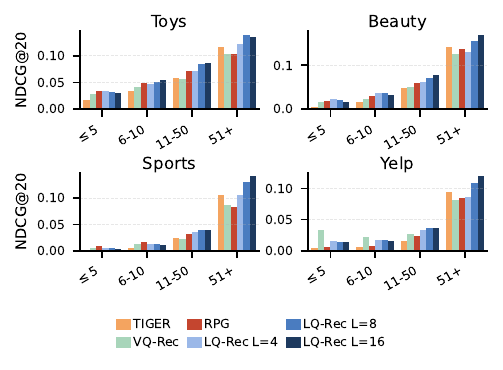}
\vspace{-1em}
\caption{Results stratified by training-set frequency of the target item. Buckets are $\le 5$, $6$--$10$, $11$--$50$, and
  $51+$.}
\label{fig:gen-freq}
\end{figure}

Two cross-dataset patterns emerge (Fig.~\ref{fig:gen-freq}).
(1) LQ-Rec at the larger $L$'s dominates the high-frequency bucket ($11+$) on every dataset; this is where joint codebook--ranking training pays off most, since the codebook receives enough signal per item to be fully refined.
(2) TIGER is the worst method in both cold buckets ($\le 10$) on every dataset, despite being the strongest baseline on the Gen side of Table~\ref{tab:gen-memgen}: this suggests that TIGER's AR decoder fails to generalize over codes assigned to cold items.
In the cold regime, no single content-aware baseline wins across all four datasets (LQ-Rec$_4$ on Toys and Beauty, RPG on Sports, VQ-Rec on Yelp): content priors generalize beyond ID memorization, but no single prior dominates---a genuine open question for future content-encoder design.

\subsection{Ablation and sensitivity (RQ4)}
\label{sec:ablation}

We consider the following variants to isolate the contribution of each component of LQ-Rec:
\textbf{T\textsubscript{posthoc}}: Stage~1 training with $\alpha=0$ (i.e., no quantization-aware training) followed by
the post-hoc residual quantization;
\textbf{T\textsubscript{stage1}}: the Stage~1 S2H-QA endpoint
with its codebook used as-is (i.e., no Stage~2);
\textbf{T\textsubscript{naive\,S2}}: Stage~1 anchor followed by Stage~2 fine-tuning at single prefix length $L = M$ (i.e., no prefix-MRL averaging).
\textbf{T\textsubscript{chain}}: the full LQ-Rec
(Section~\ref{sec:method}).
All four share $M = 16$, $K = 256$.
We additionally report
\textbf{T\textsubscript{noQ}}, the $L$-independent continuous-scoring reference obtained by scoring with the same Stage~1 S2H-QA encoder as \textbf{T\textsubscript{stage1}}\,/\,\textbf{T\textsubscript{naive\,S2}}\,/\,\textbf{T\textsubscript{chain}} but without any item quantization.

\begin{table}[t]
\centering
\caption{Ablation Study of Quantization and Training Components ($M=16$, $K=256$, R@10).}
\label{tab:ablation-tier}
\footnotesize
\setlength{\tabcolsep}{3pt}
\begin{tabular}{l r c c c c c}
  \toprule
  Dataset & $L$
    & T\textsubscript{posthoc}
    & T\textsubscript{stage1}
    & T\textsubscript{naive\,S2}
    & T\textsubscript{chain}
    & T\textsubscript{noQ} \\
  \midrule
  \multirow{3}{*}{Toys}
    & $4$  & 0.0810 & 0.0838 & 0.0822 & 0.0860 & \multirow{3}{*}{0.0987} \\
    & $8$  & 0.0903 & 0.0966 & 0.0966 & 0.0966 & \\
    & $16$ & 0.0976 & 0.1026 & 0.1032 & 0.1037 & \\
  \midrule
  \multirow{3}{*}{Beauty}
    & $4$  & 0.0813 & 0.0815 & 0.0806 & 0.0874 & \multirow{3}{*}{0.0917} \\
    & $8$  & 0.0912 & 0.0959 & 0.0951 & 0.0972 & \\
    & $16$ & 0.0952 & 0.0997 & 0.1001 & 0.0995 & \\
  \midrule
  \multirow{3}{*}{Sports}
    & $4$  & 0.0508 & 0.0523 & 0.0517 & 0.0511 & \multirow{3}{*}{0.0605} \\
    & $8$  & 0.0552 & 0.0585 & 0.0591 & 0.0585 & \\
    & $16$ & 0.0600 & 0.0633 & 0.0631 & 0.0613 & \\
  \midrule
  \multirow{3}{*}{Yelp}
    & $4$  & 0.0484 & 0.0482 & 0.0483 & 0.0538 & \multirow{3}{*}{0.0533} \\
    & $8$  & 0.0569 & 0.0592 & 0.0586 & 0.0611 & \\
    & $16$ & 0.0622 & 0.0648 & 0.0645 & 0.0640 & \\
  \bottomrule
\end{tabular}
\end{table}

Four pairwise comparisons against T\textsubscript{chain} isolate the
contribution of each component:\\
\textbf{T\textsubscript{posthoc}$\to$T\textsubscript{chain}}:
the full method dominates the natural ``train a content-based
ranker, then fit a separately optimized residual quantizer''
procedure on every dataset at $L = M$, confirming that training
the codebook jointly with the ranking objective is load-bearing
(capability~(i)).
\\
\textbf{T\textsubscript{stage1}$\to$T\textsubscript{chain}}:
Stage~2 contributes mainly at short prefix lengths; at $L = M$ the
chain is within run-to-run noise of T\textsubscript{stage1}, so the
Stage~2 gain is not extra optimization on the full-length
objective.
\\
\textbf{T\textsubscript{naive\,S2}$\to$T\textsubscript{chain}}:
prefix-MRL Stage~2 unlocks anytime serving --- materially larger
$L = 4$ R@10 at near-zero $L = M$ cost
(T\textsubscript{naive\,S2} sits at most $0.002$ R@10 above
T\textsubscript{chain} at $L = M$ on three of four datasets; Toys
reverses the sign).
\\
\textbf{T\textsubscript{noQ}$\to$T\textsubscript{chain}}:
the chain at $L = M$ \emph{exceeds} the unquantized score of its
own encoder on every dataset: under the quantization-aware
objective the encoder geometry has become specialized for the
codebook, so scoring against the quantized residual centroids
beats scoring against the continuous embedding the encoder was
nominally trained on (Appendix~\ref{app:rankloss-taskopt}).

In addition, we conducted parameter sensitivity analyses. 
We confirmed that both the $\alpha$-schedule ratios
and the codebook learning-rate multiplier were not sensitive (see Appendix~\ref{app:add-alpha}).

\subsection{Latency and scalability (RQ5)}
\label{sec:latency}

We compare inference cost across three representative semantic-ID
decoding paradigms: \emph{late quantization} (LQ-Rec),
\emph{early quantization with autoregressive decoding} (TIGER), and
\emph{graph-based decoding} (RPG). 

\subsubsection{Query latency}
\label{sec:latency-wallclock}

\begin{figure}[t]
\centering
\includegraphics[width=0.88\linewidth]{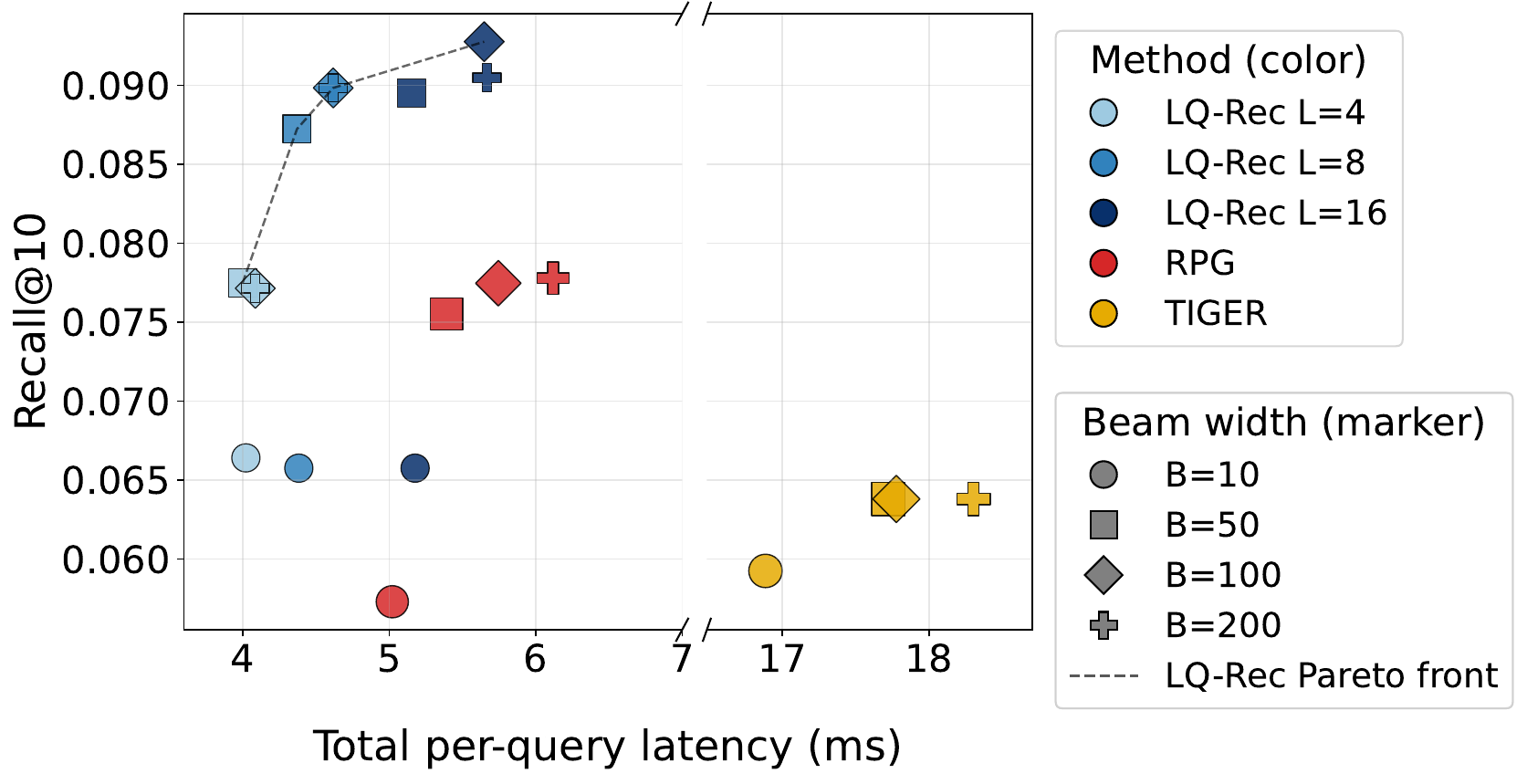}
\caption{Single-query latency $\times$ R@10, mean across Toys,
Beauty, and Sports.}
\label{fig:latency-pareto}
\vspace{-0.4em}
\end{figure}

Figure~\ref{fig:latency-pareto} plots per-query latency against
Recall@10 averaged over Toys, Beauty, and Sports, sweeping the beam
width $b \in \{10, 50, 100, 200\}$. 
LQ-Rec latencies include a
per-dataset SBERT batch-encode offset: LQ-Rec runs the text
encoder on the user's history at query time, whereas TIGER and RPG
look up precomputed semantic IDs.

LQ-Rec dominates the Pareto front: it is both the fastest and the
most accurate, and sweeping $L$ traces out a clean recall--latency
trade-off that lifts the entire front.  RPG sits slightly below due
to the extra cost of per-step graph
traversal.  TIGER is slower despite $M{=}4$, for the
structural reason analyzed next.  Beam width saturates by
$b \approx 100$ on all three methods, so the choice of $b$ does
not separate them.

\subsubsection{Why TIGER is slow}
\label{sec:latency-asymptotic}

The gap is structural, not a constant-factor implementation
difference.
TIGER's autoregressive decoder issues $M{=}4$ depth-wise
transformer forwards on top of the encoder, run strictly in sequence.
TIGER pays this $3\times$ wall-clock penalty despite using a
shorter code ($M{=}4$ vs.\ LQ-Rec $L{\in}\{8,16\}$): the serial
$M$-step chain, each forward carrying a $b$-fold larger batch,
dominates the budget (Table~\ref{tab:complexity}, right column).
LQ-Rec and RPG instead run the encoder once, independent of semantic ID length.

\begin{table}[t]
\centering
\caption{Memory and transformer-forward count per query.}
\label{tab:complexity}
\footnotesize
\setlength{\tabcolsep}{4pt}
\begin{tabular}{lll}
\toprule
Method                          & Catalog memory          & \# GPU fwd / q \\
\midrule
SASRec~\cite{sasrec}            & $O(Nd)$                 & $1$            \\
VQ-Rec~\cite{vqrec}             & $O(NM + MKd)$           & $1$            \\
TIGER~\cite{tiger}              & $O(NM + MKd)$           & $1 + M$ ($\times b$ batch) \\
RPG~\cite{rpg}                  & $O(N(M{+}k) + MKd)$     & $1$            \\
\textbf{LQ-Rec (ours)}          & $O(NM + MKd)$           & $1$            \\
\bottomrule
\end{tabular}
\\[3pt]
{\scriptsize $N$: catalog size; $d$: encoder hidden dim;
$M$: per-item code length (residual codebooks for TIGER/LQ-Rec,
OPQ subspaces for RPG); $K$: per-codebook size; $b$: beam width;
$k$: RPG per-item graph out-degree~\cite{rpg}.}
\end{table}

\subsubsection{Scalability to catalog size $N$}
\label{sec:latency-N}

The Amazon benchmarks have $N \lesssim 2 \times 10^4$
items. We probe larger $N$ by re-benching the three paradigms on
Sports ($N_{\text{native}}{=}18{,}357$) with additional item injection
up to $N = 5 \times 10^5$.
We benchmark latency and catalog memory below.

\begin{figure}[t]
\centering
\includegraphics[width=\linewidth]{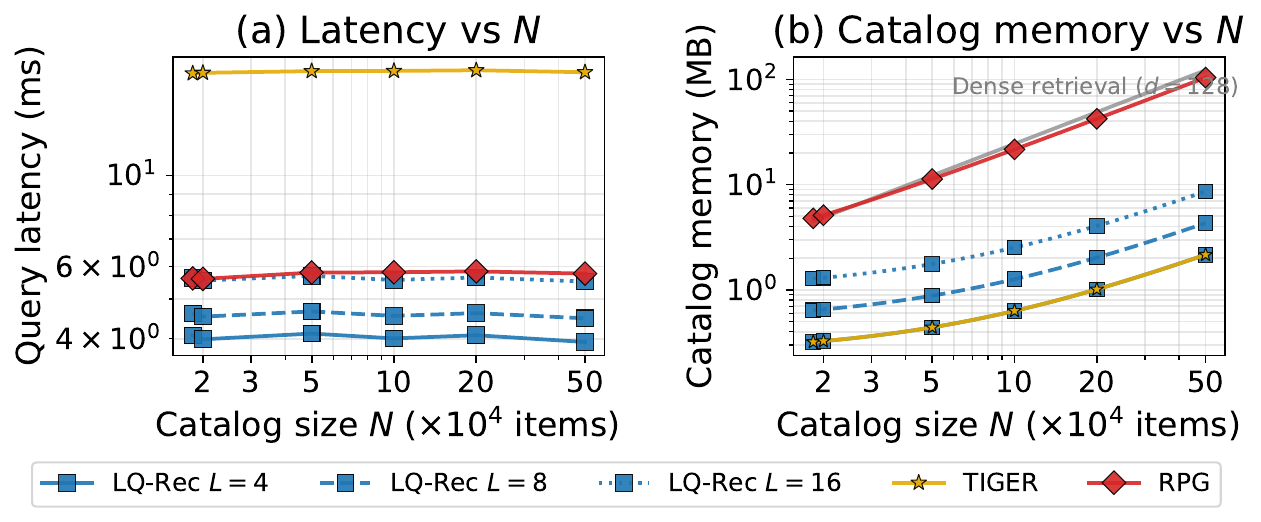}
\vspace{-1.7em}
\caption{Scalability against item catalog size $N$ on Sports. 
(a) Query latency,
(b) Catalog memory under common configurations.}
\label{fig:nsweep-3methods}
\end{figure}

Per-query latency (Fig.~\ref{fig:nsweep-3methods}a)
is flat in $N$ because the retrieval algorithms do not depend on $N$
(fixed-cost transformer forwards dominate).
The wall-clock ordering of
Sec.~\ref{sec:latency-wallclock} therefore extends unchanged to a
$27\times$ larger catalog.

We measure memory for item catalog storage under common assumptions
(Fig.~\ref{fig:nsweep-3methods}b; $K{=}256$ codes packed as 1\,B,
$d{=}128$, FP16 codebook, int32 RPG adjacency).
TIGER ($M{=}4$) and LQ-Rec $L{=}4$ overlap;
LQ-Rec ($L{=}8, 16$) trade memory for quality;
RPG is the largest because the
$O(Nk)$ graph adjacency dominates ($\sim 100$~MB).
LQ-Rec at $L{=}16$ is roughly $12\times$ lighter than RPG at the same $N$.

\subsection{Case study: semantic hierarchy of the codebook (RQ6)}
\label{sec:case-study}

LQ-Rec fits its codebook from a next-item ranking loss alone,
with no reconstruction objective on item content like TIGER~\cite{tiger}.
Despite this, a TIGER-like hierarchical structure still emerges.
Table~\ref{tab:prefix-purity} reports the size-weighted
dominant-leaf-category purity of $L$-prefix groups (i.e., the
fraction of items in each group that share the group's most-common leaf category, averaged over groups weighted by size), where each item's category label is taken as the leaf (deepest node) of its first
Amazon category path; purity rises monotonically with $L$ and reaches
$100\%$ at $L{=}16$ on every dataset.

\begin{table}[t]
\centering
\caption{Size-weighted dominant-leaf-category purity (\%) of LQ-Rec.}
\label{tab:prefix-purity}
\footnotesize
\setlength{\tabcolsep}{6pt}
\begin{tabular}{lcccccc}
  \toprule
  Prefix length $L$ & $1$ & $2$ & $3$ & $4$ & $8$ & $16$ \\
  \midrule
  Toys    & 39 & 62 & 82 & 92 & 100 & 100 \\
  Beauty  & 31 & 57 & 77 & 89 &  95 & 100 \\
  Sports  & 24 & 50 & 70 & 87 &  98 & 100 \\
  \bottomrule
\end{tabular}
\end{table}

Table~\ref{tab:case-study-examples} shows the 
qualitative patterns: a shared first-stage code already isolates a
brand or category supercluster, and the second stage refines it
into distinct sub-categories.

\definecolor{headA}{HTML}{C0392B}
\definecolor{headB}{HTML}{2874A6}
\definecolor{headC}{HTML}{1E8449}
\begin{table}[t]
\centering
\caption{Sibling $L{=}2$ prefix groups: shared first-stage code
(colored) splits into distinct sub-categories at the second stage.}
\label{tab:case-study-examples}
\footnotesize
\setlength{\tabcolsep}{4pt}
\renewcommand{\arraystretch}{1.15}
\begin{tabular}{@{}l l p{0.60\columnwidth}@{}}
  \toprule
  Dataset & Prefix & Sub-cluster contents (size, sample items) \\
  \midrule
  Toys   & $[\textcolor{headA}{\mathbf{237}}, 105]$ & \textbf{Floor puzzles} ($n{=}5$): Melissa \& Doug Farm Maze, Pirate's Bounty, Fairy Tale Castle \\
         & $[\textcolor{headA}{\mathbf{237}}, 4]$   & \textbf{Jigsaw puzzles} ($n{=}3$): Melissa \& Doug Playful Pets, African Plains \\
  \midrule
  Beauty & $[\textcolor{headB}{\mathbf{18}}, 198]$  & \textbf{Lipstick} ($n{=}5$): Wet n Wild Mega Last Lip Color (\#919B, \#966, \ldots) \\
         & $[\textcolor{headB}{\mathbf{18}}, 96]$   & \textbf{Eye shadow} ($n{=}4$): Wet n Wild Coloricon Eye Shadow Trio (\#738, \ldots) \\
  \midrule
  Sports & $[\textcolor{headC}{\mathbf{189}}, 148]$ & \textbf{Bike shop tools} ($n{=}9$): IceToolz Steel Tire Levers, Avenir Pro Spoke Wrench \\
         & $[\textcolor{headC}{\mathbf{189}}, 253]$ & \textbf{Bike tubes} ($n{=}3$): Sunlite Thorn Resistant Bicycle Tubes \\
  \bottomrule
\end{tabular}
\end{table}

\section{Conclusion}
\label{sec:conclusion}

We introduced \textbf{Late Quantization (LQ)}, a non-autoregressive
paradigm that confines discretization to the item side and keeps the encoder and scoring rule in continuous space.
This single structural move, a choice of \emph{where} the quantizer sits rather than \emph{how} it is designed, removes the AR-GR coupling that caps the semantic-ID length, and yields structural guarantees rather than mere empirical wins: the retrieval error decomposes additively
(Proposition~\ref{prop:error-decomp}), the encoder objective is
quantizer-invariant so encoder and codebook can be improved nearly independently (Proposition~\ref{prop:separable}),
and the quantization-induced ranking drop is bounded by per-item errors and pairwise margins (Theorem~\ref{thm:pairwise}).
We instantiated the framework as \textbf{LQ-Rec} with a two-stage chain (S2H-QA + Prefix-MQA), whose Prefix-MQA stage further collapses GR's ``one model per $M$'' regime into a single anytime checkpoint.
On four public benchmarks, LQ-Rec delivers $M$-scalability,
single-checkpoint anytime inference, and generalization to unseen patterns, the first
non-autoregressive semantic-ID method to do so, while
outperforming every baseline. %

\bibliographystyle{IEEEtran}
\bibliography{references}

\appendix

\subsection{Proofs of Theorem~\ref{thm:pairwise} and Corollary~\ref{cor:recall}}
\label{app:proofs}

\noindent\textbf{Theorem~\ref{thm:pairwise} proof}:  First, we establish
$|\hat{s}_{ij} - s_{ij}| \le \norm{u}(\varepsilon_i + \varepsilon_j)$.
\begin{align*}
  \hat{s}_{ij} - s_{ij}
  &= \inner{u}{\hat{e}_i - \hat{e}_j} - \inner{u}{e_i - e_j} \\
  &= \inner{u}{(\hat{e}_i - e_i) - (\hat{e}_j - e_j)}
  = \inner{u}{\delta_i - \delta_j},
\end{align*}
so by the triangle inequality and Cauchy--Schwarz
\(
  |\hat{s}_{ij} - s_{ij}|
  \le |\inner{u}{\delta_i}| + |\inner{u}{\delta_j}|
  \le \norm{u}\,(\varepsilon_i + \varepsilon_j).
\)
If $s_{ij} > 0$, a violation $\hat{s}_{ij} \le 0$ requires
$s_{ij} \le s_{ij} - \hat{s}_{ij} \le |s_{ij} - \hat{s}_{ij}|
\le \norm{u}\,(\varepsilon_i + \varepsilon_j)$
by the score perturbation bound just established.
Dividing by $\norm{u} > 0$ gives
$m_{ij} = s_{ij}/\norm{u} \le \varepsilon_i + \varepsilon_j$.
Taking the contrapositive yields the claim;
independence from $\norm{u}$ is immediate since
$m_{ij} = \inner{\udir}{e_i - e_j}$ depends only on the direction.
A violation can occur only when
$m_{ij} \le \varepsilon_i + \varepsilon_j$, so
$\mathrm{VR}(u) \le
 \Prob_{(i,j):\,s_{ij}>0}\!\bigl(m_{ij} < \varepsilon_i + \varepsilon_j\bigr)$.

\noindent\textbf{Corollary~\ref{cor:recall} proof.}
Suppose $i^*$ has continuous rank $r \le K$.
The quantized rank of $i^*$ is
$\hat{r} = 1 + |\{j \ne i^* : \hat{s}(j) > \hat{s}(i^*)\}|$.
For $\hat{r} > K$, at least one item $j$ with continuous rank $> K$ must
satisfy $\hat{s}(j) > \hat{s}(i^*)$, i.e.\ the pair $(i^*, j)$ is
violated.
By Theorem~\ref{thm:pairwise}(a), this requires
$m_{i^* j} \le \varepsilon_{i^*} + \varepsilon_j$, i.e.\
$\tilde{\gamma}_{i^*} \le 0$.
Taking complementary probabilities and averaging over $(u, i^*)$
gives the population bound.

\subsection{Tightness of the Pairwise Ordering bound}
\label{app:rankloss-tightness}

The Cauchy--Schwarz step
$|\inner{u}{\delta_i - \delta_j}| \le \norm{u}(\varepsilon_i + \varepsilon_j)$
is tight only when $\delta_i - \delta_j \parallel u$ (parallel).
Empirically, this perturbation is far below the worst-case bound (e.g., median $4.5\%$ of $\|u\|(\varepsilon_i+\varepsilon_j)$ on Toys, with no test pairs violating the bound). Since it takes either sign, optimizing this directional term directly beats one attacking $\varepsilon_i$.

This is what $\mathcal{L}_{\mathrm{quant}}$ provides.
With logit $\hat{s}_j = \inner{u}{\hat{e}_j}$ and temperature $\tau$,
the cross-entropy gradient
$\partial \mathcal{L}_{\mathrm{CE}} / \partial \hat{s}_j = p_j / \tau$
(where $p_j \propto \exp(\hat{s}_j/\tau)$) concentrates on the
top-ranked negative---the most severe potential pairwise violator.
Decomposing
$\hat{s}_{i^* j} = s_{i^* j} + \inner{u}{\delta_{i^*} - \delta_j}$,
this gradient acts on the second term directly---the exact quantity
that Theorem~\ref{thm:pairwise} bounds via Cauchy--Schwarz---and
pushes it in the favorable sign rather than driving its norm to zero.
By contrast, $\mathcal{L}_{\mathrm{commit}}$ acts only on
$\varepsilon_i$ and operates within the bound;
$\mathcal{L}_{\mathrm{quant}}$ exploits the slack the bound leaves
behind.

\subsection{Approximation optimal vs.\ task optimal}
\label{app:rankloss-taskopt}
Here we collect the optimization analysis that motivates two design
choices made in the main text: LQ-Rec's Stage~2 omits an MSE
pull-back term $\lambda \norm{e_i - \hat{e}_i}^2$
(i.e.~$\lambda = 0$), and Stage~1 keeps the commitment regularizer
$\beta\,\mathcal{L}_{\mathrm{commit}}$.

Two distinct notions of codebook optimality emerge under finite
codebook capacity.
\textbf{Approximation optimal}: $\hat{e}_i \approx e_i$
(minimise $\varepsilon_i$).
\textbf{Task optimal}: $\hat{e}_i$ maximizes Recall@$K$ under the
constraint that each $\hat{e}_i$ lies in the codebook reconstruction
set.
$\mathcal{L}_{\mathrm{CE}}$ on $\hat{s}_q$ drives codebooks toward
task optimality, possibly increasing $\varepsilon_i$;
the MSE pull-back $\lambda \norm{e_i - \hat{e}_i}^2$ drives them
toward approximation optimality, constraining CE.

In Stage~1, by contrast, the encoder is trainable, the commitment
loss $\norm{e_i - \sg(q_i)}^2$ has the codebook detached, and its
gradient flows only to the encoder.
$\mathcal{L}_{\mathrm{commit}}$ stabilizes the encoder against drift
from the codebook reconstruction set under STE; the codebook
remains free to move toward task-optimal placement under
$\mathcal{L}_{\mathrm{quant}}$.

\subsection{Sensitivity analysis}
\label{app:add-alpha}

Let $(\rho_A, \rho_P) = (T_A, T_P)/T_{\max}$.
Sweeping eight cells of $(\rho_A, \rho_P)$ with
$\rho_A \in [0.10, 0.55]$, $\rho_P \in [0.30, 0.70]$,
$\rho_A \le \rho_P$ (anchor $(0.25, 0.50)$ at R@10$=0.0869$), the 3-dataset mean R@10
stays in $[0.0863, 0.0876]$ across all eight cells.

Sweeping $\mathrm{mult} \in \{1, 2, 5\}$ on the same three datasets
(anchor $\mathrm{mult} = 2$ at R@10$=0.0876$), the
3-dataset mean R@10 lies in $[0.0867, 0.0876]$.

\subsection{TIGER scaling with deeper RQ codebooks}
\label{app:tiger-depth}

We evaluated TIGER with longer semantic IDs.
Test R@10 was
$0.0718/\textbf{0.0084}$ on toys ($M{=}4/8$) and
$0.0432/\textbf{0.0107}$ on sports ($M{=}4/8$), showing significant drops in $M$;
on beauty, where R@10 was more stable, we also ran $M{=}16$,
giving $0.0668/0.0677/\textbf{0.0487}$ for $M{=}4/8/16$.
This drop at deeper $M$ is consistent with Hou et al.~\cite{rpg} (their Table~7) and shows  that TIGER is hard to stably generalize.

\end{document}